\documentclass{article}

\usepackage{arxiv}

\usepackage[utf8]{inputenc} 
\usepackage[T1]{fontenc}    
\usepackage{hyperref}       
\usepackage{url}            
\usepackage{booktabs}       
\usepackage{amsfonts}       
\usepackage{nicefrac}       
\usepackage{microtype}      
\usepackage{lipsum}		
\usepackage{graphicx}
\usepackage{natbib}
\usepackage{doi}

\usepackage{amsmath}
\DeclareMathOperator*{\argmax}{argmax}
\usepackage[linesnumbered,ruled]{algorithm2e}

\title{Identifying blockchain-based cryptocurrency accounts using investment portfolios}

\author{ \href{https://aminrd.github.io}{\includegraphics[scale=0.06]{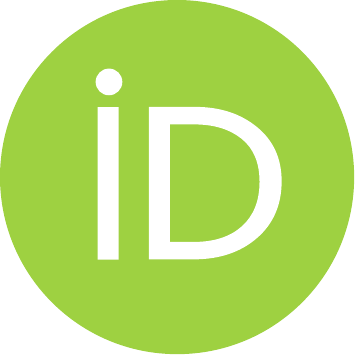}\hspace{1mm}Amin Aghaee}
	\\ Vancouver, BC, Canada\\
	\texttt{https://aminrd.github.io} \\
}

\renewcommand{\undertitle}{A Case Study}

\hypersetup{
pdftitle={Identifying blockchain-based cryptocurrency accounts using investment portfolios},
pdfauthor={Amin Aghaee},
pdfkeywords={Blockchain, Cryptocurrency, Identification, Social Media, Algorithms},
}

\begin{document}
\maketitle

\begin{abstract}
	Cryptocurrencies based on decentralized systems, especially blockchain, are gaining popularity more than ever. Freedom advocates hail blockchain technology as a breakthrough in digital privacy and internet anonymity. Unfortunately, after recent studies conducted, it may come as a surprise that the transactions are, in fact, not always anonymous. In this short paper, the possibility of identifying a user's accounts in different cryptocurrencies given the user's portfolio of investment gained from social media is investigated. In this study, the generic elements of blockchain systems are briefly studied. In section \ref{sec:blocksim}, BlockSim which is a tool for simulating transactions, and an algorithm for answering this question is introduced.
\end{abstract}

\keywords{Blockchain\and Cryptocurrency\and Identification\and Social Media \and Approximate Algorithms}

\section{Introduction}
A cryptocurrency (also known as crypto) is a digital currency in which transactions are verified and records maintained by a decentralized system using cryptography, rather than by a centralized authority. Cryptography provides a mechanism for securely encoding the rules of a cryptocurrency system in the system itself. \cite{cryptobook} Many recent popular cryptocurrencies (e.g. Bitcoin, Ethereum, etc.) are decentralized networks based on blockchain technology.

In recent years, the growth of Bitcoin, Ethereum, and other cryptocurrencies is a phenomenon that has attracted unprecedented attention. \cite{caporale2018persistence} For years, markets and currencies have been controlled by centralized systems such as governments and central banks. However, the usage of cryptocurrencies keeps evolving day by day due to some benefits these new decentralized currencies suggest, such as privacy, security, access, and efficiency. 

In most blockchain systems, especially in cryptocurrencies, all transactions, the wallet addresses involved are recorded on a public chain and are available for everyone. Although wallet addresses alone do not disclose identifiable details, they do provide some information for further investigation. Revealing some contact-related information and user identifications has gained attraction in recent years \cite{chang2018improving}. Alqassem et al. for example studied ways for suggesting some techniques for Bitcoin Data analysis \cite{alqassem2018anti}.

The effect of social media on almost everything is not deniable these days. The cryptocurrency market is controlled by many factors including social media \cite{aggarwal2019understanding}. As an example, many people post the latest trends related to cryptocurrencies on Reddit, and share their portfolio of investment on social media. In this paper, the question of whether we can disclose users' accounts using their portfolio of investment in multiple cryptocurrencies is investigated.

\section{Background And Problem Definition}
\label{sec:background}
In this section, some public characteristics of a blockchain system are briefly described. A comprehensive study of blockchain systems is well written in \cite{bashir2020mastering}. Each blockchain system consists of some fundamental elements. Some of the generic elements in each blockchain system, or more specifically in a cryptocurrency blockchain system which is related to this work are \cite{bashir2017mastering}: 
\begin{itemize}
	\item Addresses: Unique identifiers being assigned to accounts
	\item Transactions: Information about transferring some value from a source address to a destination address
	\item Blocks: Multiple transaction information, the previous and the next block, and some other meta information are stored in a block
	\item Distributed consensus: Enables a blockchain system to have a single trusted history shared between all people and nodes in the network
\end{itemize}

When user $A$ wants to transfer some value $v$ from his account $a_A$ to another account $a_B$ which belongs to user $B$, the following steps happen \cite{bashir2017mastering}: 
\begin{enumerate}
    \item Sender $A$ signs the transaction using his/ her private key
    \item The transaction information is broadcasted to the network using a specific algorithm called flooding
    \item Miner nodes add the information about this transaction to their block
    \item After a couple of transactions happened, a miner who can provide a proof of work (PoW), can add a reward value $v_r$ to his/ her account $a_m$ and broadcasts the solved problem to the network
    \item Every other node and miner in the network verify and add the recently solved block into their block. Thus, the information about the transaction between accounts $a_A, a_B$ which is inside the recently-solved block is also broadcasted
    \item Finally, the value starts to be appeared in $a_B$ after between three to six confirmations
\end{enumerate}

When it comes to privacy protection in a distributed database, blockchain systems have significant advantages \cite{wang2020survey, zyskind2015decentralizing}. The history of all transactions that happened in a public blockchain system is considered public information and is accessible to everyone. For example, the history of all transactions in Bitcoin from the first transaction in January 2009 is publicly available. Nonetheless, the public address of the accounts is stored in the blocks rather than user identities. For instance, you can have an account in Bitcoin which has a public address that only you and those who want to send money to you know whose this address belongs to. For other people, this public address may seem like a series of random numbers.

Nonetheless, the privacy and confidentiality of users is still a debatable topic that has been studied in recent years (\cite{menegalli2021proving, conti2018survey, zaghloul2020bitcoin, androulaki2013evaluating}). In this paper, a case is studied in which the following question is investigated: Are we able to identify users' accounts in different cryptocurrencies having their portfolio of investment? Each user can have one or multiple accounts. In this paper, we study the simple case of this problem by limiting the number of accounts users can have in a single cryptocurrency to one. In other words, every user can hold at most one account in cryptocurrencies such as Bitcoin, Ethereum, etc. 

In order to define the problem officially, let say we have a set of $m$ cryptocurrencies $C = \{C_1, C_2, ..., C_m\}$. User $A$ has one account in each of these cryptocurrencies. $U$'s portfolio of investment is: 
$$
P = \{\alpha_1, \alpha_2, ..., \alpha_m \}, \forall i \leq i \leq m: 0 \leq \alpha_i \leq 1, \sum_{i=1}^{m} \alpha_i = 1
$$

That means if $v$ is the total investment of user $U$ in all cryptocurrencies, $\alpha_1 \times v$ is invested in currency $C_1$, $\alpha_2 \times v$ is invested in currency $C_2$ and etc. User $U$ holds these amounts in accounts $A_U = \{a_u^1, a_u^2, ..., a_u^m\}$ where $\forall i: a_u^i \in C_i$. Finding $A_U$ given $P$ is studied in this article. 

\section{BlockSim}
\label{sec:blocksim}
One of the challenges of this problem in a real scenario is that we do not know the account owners of different cryptocurrencies. BlockSim \footnote{https://github.com/aminrd/BlockSim}, is a Python package for simulating the transactions of multiple cryptocurrencies. In BlockSim, you can define: 

\begin{itemize}
    \item User acquisition rate, which is the rate of adding new users to a cryptocurrency over time
    \item Number of turns for simulation
    \item Number of cryptocurrencies in your system
    \item Number of transactions in each turn in a turn
    \item Miner rate for each cryptocurrency, which specifies the transaction fee described in \ref{sec:background}
\end{itemize}

Once these variables are defined, the simulator starts storing the transaction information into a database, in which records of all simulated users, their accounts, and all transactions are stored. Note that in a real scenario, since all transactions of many cryptocurrencies are public, it is possible to go through the transactions one by one and recreate the database. For example, let say we have the investment portfolio of a user in Bitcoin, Ethereum, and Cardano at time $t$. We can start from time $0$ and create three different databases of $\{C_{Bitcoin}, C_{Ethereum}, C_{Cardano}\}$ each of which contains multiple accounts. To speed up our process of finding $A_u$, we drop all accounts holding zero balances. It is important to note that, when we refer to balance in this case, we refer to balances all converted into a common currency (either a fiat currency or a cryptocurrency) at time $t$ so that we can compare the value of an account in $C_{Bitcoin}$ to the value of another account in $C_{Ethereum}$. 

Let say a possible answer to the problem given investment portfolio $P = \{\alpha_1, \alpha_2, ..., \alpha_m \}$ is $A_U = \{a_u^1, a_u^2, ..., a_u^m\}$. In order to evaluate this answer and compare it to other answers, we need to define a score. The following score is proposed: 

$$
score = m - \sum_{a_i \in A_U} | \alpha_i - \frac{a_i \text{.balance}}{\sum_{a \in A_U}a\text{.balance}} | 
$$

A pseudo-code for the algorithm of finding accounts that respect the investment portfolio is in algorithm \ref{alg:finder}. The method starts with creating the databases of all cryptocurrencies from time $0$ to time $t$ sorted by the number of non-zero-balance accounts. Thus, the first crypto database on the list is $C^t_{sorted}(1)$ that has the least number of accounts. All accounts in each crypto database are also sorted by balances. Then, for each account $a \in C^t_{sorted}(1)$, other potential accounts in other crypto databases are looked for. Note that, for each account $a \in C^t_1$, we are looking for an account $b \in C^t_i$ so that $\frac{a\text{.balance}}{b\text{.balance}} = \frac{\alpha_1}{\alpha_i}$. Thus, we know what is the target balance in other crypto databases to look for. The method for finding one or multiple accounts in a crypto database is defined in \textit{BinaryFind} \ref{alg:bfind}.

\begin{algorithm}
\label{alg:finder}
    \SetKwInOut{Input}{Input}
    \SetKwInOut{Output}{Output}
    \Input{Portfolio $P = \{\alpha_1, \alpha_2, ..., \alpha_m \}$, Cryptocurrencies $C = \{C_1, C_2, ..., C_m\}$, Time $t$, Score Threshold $S_t$}
    \Output{An answer $A^* = \{a^*_1, a^*_2, ..., a^*_m\}$}
    
    $C^t \gets $ All cryptocurrencies in $C$ sorted by their number of accounts at time $t$\;
    
    $C^t_{sorted} \gets \{sort(C^t_1), sort(C^t_2), ..., sort(C^t_m)\}$ by account balances at time $t$\;
    
    $Answers \gets$ an empty set $\{\}$ \;
    
    \ForEach{$a \in C^t_{sorted}(1)$.accounts}{
        
        $A_1 \gets \{a\}$ \;
        \For{$i \in \{2,3,...,m\} $ }{
            $A_{i} \gets$ BinaryFind($C^t_{sorted}(i)$.accounts, $a$.balance $\times \frac{\alpha_i}{\alpha_1}$)
        }
        \ForEach{$ans \in \{A_1 \times A_2 \times ... \times A_m \}$}{
            \If{ score($ans$) $\geq S_t$}{
                $Answers$.add($ans$) \;
            }
        }
    }
    
    \Return $\argmax_{A^*} score(A^*)$ where $A^* \in Answers$\;

    \caption{Algorithm for finding potential accounts satisfying a portfolio holder}
\end{algorithm}

BinaryFind \ref{alg:bfind} algorithm is inspired by the binary search method with some adjustments. The run-time complexity of binary search algorithm on a sorted list having $n$ elements is $\mathcal{O}(n \log n)$. If the algorithm finds the target balance in the list of accounts, it returns all accounts having that balance. If no accounts are found, then at least two accounts as the lower bound and higher bounds of the target balance would be returned. For instance, if a list of accounts is:
$$
\{(id, balance)'s\} = \{(1, 1.23), (2, 3.78), (3, 6.0), (4, 6.0), (5, 7.13), (6, 8.2), (7, 12.6) \}
$$

given a target balance $b=6.6$, the output would be $\{(3, 6.0), (4, 6.0)\}$ and given a target balance $b=7.99$, the output would be $\{ (5, 7.13), (6, 8.2)\}$.

\begin{algorithm}
\label{alg:bfind}
    \SetKwInOut{Input}{Input}
    \SetKwInOut{Output}{Output}
    \Input{Target balance $b$, Accounts $\{a_1, a_2, ..., a_n\}$ sorted by balances}
    \Output{List of $p$ answers $\{a^*_1, a^*_2, ..., a^*_p\}$}
    
    \If{$a_1\text{.balance} > b$ or $a_n\text{.balance} < b$}{
        \Return $\{\}$\;
    }
    
    $left, right \gets 1, n$ \;
    
    \While{$right > left$}{
        $mid \gets \frac{left + right}{2}$ \;
        \If{$a_{mid}\text{.balance} == b$}{
            $ans \gets \{a_{mid-l^*}, ..., a_{mid}, ..., a_{mid+r^*}\}$ where $\forall a \in ans: a$.balance $==b$\;
            \Return $ans$
        }
        \ElseIf{$a_{mid}\text{.balance} > b$}{
            \If{$right == mid$}{
                \Return $\{a_{left}, ..., a_{right}\}$ \;
            }\Else{
                $right \gets mid$\;
            }
        }
        \Else{
            \If{$left == mid$}{
                \Return $\{a_{left}, ..., a_{right}\}$ \;
            }\Else{
                $left \gets mid$\;
            }        
        }
    }

    \caption{BinaryFind a target balance in a list of accounts sorted by balances}
\end{algorithm}

After finding all possible accounts in each crypto databases, algorithm \ref{alg:finder} creates a set of possible answers derived from the Cartesian product of all possible accounts $ANS = \{A_1 \times A_2 \times ... \times A_m \}$. The smallest size of $ANS$ is $1$, one average is $2^m$ (considering the lower and upper bounds in BinaryFind's output) and could be larger if there are many accounts in each database with exactly equal balances. Then, the score is computed for each answer in $ANS$, and all answers having scores less than $S_t$ are dropped. Finally, an answer with a maximum score is returned as the output of this algorithm. A reason for using the score threshold $S_t$ is sometimes, multiple possible answers sorted by their scores might be interested rather than the best answer achieved.

To analyze the complexity of this algorithm, let say $n = \max_{c \in C} size(c)$ where size refers to the number of non-zero-balance accounts in a crypto database. The method starts by sorting the databases by their sizes in $\mathcal{O}(m \log m)$. It is clear that $m \ll n$. Thus this part of the algorithm is negligible. Then, each database is sorted by balances in $\mathcal{O}(m.n \log n)$. Then BinaryFind is called $m$ times in total complexity of $\mathcal{O}(m \log n)$. On average, if we assume that there are $2^m$ possible answers to be processed the complexity of the for loop is $\mathcal{O}(n.m.\log n + n.2^m)$. Since we know $m \ll n$, the total complexity of this algorithm would be $\mathcal{O}(m.n \log n)$.

\section{Case Study}
\begin{figure}
	\centering
	\includegraphics[width=\linewidth]{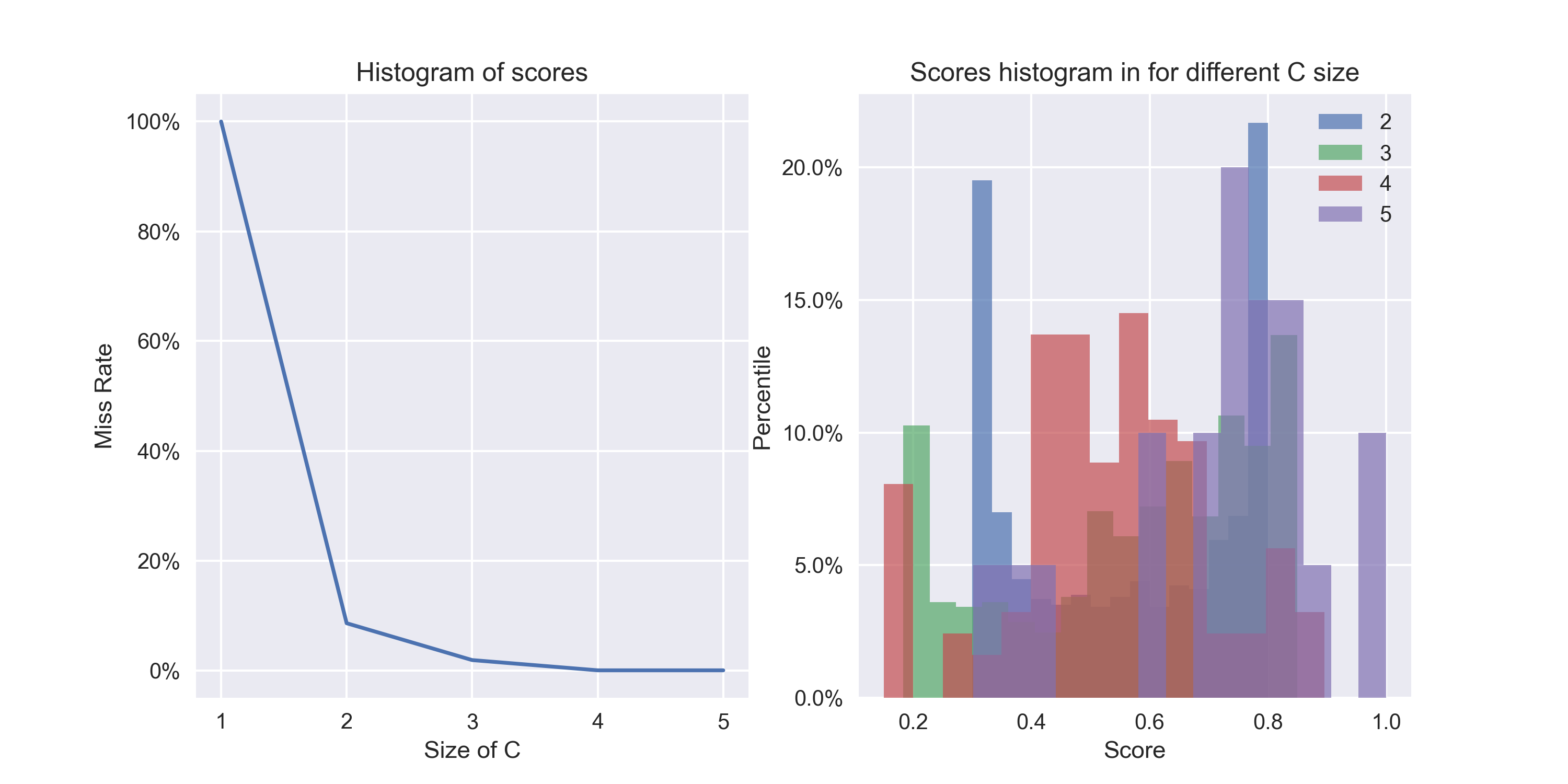}
	\caption{Missing rate vs. size of $C$ on the left, percentile histograms of normalized scores for different $C$ sizes on the right}
	\label{fig:plot}
\end{figure}

For a case study, a simulation of up to 1000 turns with five simulated cryptocurrencies was conducted. Each cryptocurrency in this simulation, has a unique linear formula of user acquisition rate at each turn $\textit{NewUsers}(t) = \beta_1 t + \beta_0$ with unique miner gifts and transaction fees. In every turn, all existing users send a random portion of his/ her balance to a random account in that currency. Summary of results gained after running \textit{Finder} method is available in figure \ref{fig:plot}. In this figure, the size of $C$ refers to the number of cryptocurrencies or $m$. In this figure, the normalized scores or $\frac{score}{m}$ are compared. The reason for using a normalized score is to map the best score obtained from the finder method between $0$ and $1$. This figure suggests an increasing number of cryptocurrency accounts in the ratio leads to a higher chance of getting a better score.

In figure \ref{fig:plot}, the missing rate of different $m$ values is plotted. Missing rate is the opposite of hitting rate which refers to cases when no solution was found. Given only one ratio of one type, the missing rate is 100 percent. There is no practical way to find an account given a single portfolio $P = \{1\}$. This plot also supports the claim that an increasing number of cryptocurrency accounts in the ratio leads to a higher chance of getting a better score.

\section{Conclusion}
In this paper, the question of identifying users' accounts in different cryptocurrencies given the investment portfolio was studied. This paper supports the fact claiming cryptocurrencies are not completely anonymous. In this study, some limitations were considered. For example, one assumption of this work is that users do not have more than 1 account in a cryptosystem, while they could have multiple accounts in different cryptosystem. Interesting problems for further studies could be to discard this limit or to find probability distributions over all accounts that refer to the confidence level rather than returning a single answer.

\bibliographystyle{unsrtnat}
\bibliography{references}  






\end{document}